\documentclass[doublecol]{epl2} 

\usepackage{amsmath}

\usepackage{amssymb}
\usepackage{bm}
\usepackage{graphicx}
\graphicspath{{figs/}} % LateX will look for figures inside of this folder. Notice that we do not a first / in front of the name of the folder.

\usepackage{xcolor}

\definecolor{LivingCoral}{RGB}{252, 118, 106}
\definecolor{OrangeRedCrayola}{RGB}{251, 90, 75} %Darkened Living Coral
\definecolor{ViridianGreen}{RGB}{0, 150, 152}
\definecolor{PacificCoast}{RGB}{91, 132, 177}

\usepackage[colorlinks=true,linkcolor=OrangeRedCrayola,
citecolor=OrangeRedCrayola,urlcolor=blue]{hyperref}

\usepackage{enumitem}
\usepackage{geometry}
\usepackage{xspace}

\usepackage{pdflscape}
\usepackage{enumerate}
\usepackage{soul} % Added to be able to strike through
\usepackage{dsfont} %To have a nice identity operator using \mathds

\newcommand{\ie}{{i.e.~}}

\newcommand{\eg}{e.g.~}

\newcommand{\etc}{\textsl{etc.\xspace}}

\let\oldsqrt\sqrt
\def\sqrt{\mathpalette\DHLhksqrt}
\def\DHLhksqrt#1#2{%
\setbox0=\hbox{$#1\oldsqrt{#2\,}$}\dimen0=\ht0
\advance\dimen0-0.2\ht0
\setbox2=\hbox{\vrule height\ht0 depth -\dimen0}%
{\box0\lower0.4pt\box2}}

\newcommand{\tr}{\mathrm{Tr}}

\newcommand{\ee}{e}

\newcommand{\boldmathsymbol}[1]{{\ensuremath{\boldsymbol{#1}}}}

\newcommand{\beq}{\begin{equation}}
\newcommand{\eeq}{\end{equation}}
\newcommand{\bea}{\begin{equation}\begin{aligned}}
\newcommand{\eea}{\end{aligned}\end{equation}}

\newlength{\wsingfig}
\newlength{\wdblefig}
\newlength{\wquadfig}
\newlength{\wtriplefig}
\newcommand{\Eq}[1]{Eq.~(\ref{#1})}
\newcommand{\Eqs}[1]{Eqs.~(\ref{#1})}
\newcommand{\Fig}[1]{Fig.~{\ref{#1}}}

\newcommand{\Refa}[1]{Ref.~{\cite{#1}}}
\newcommand{\Refs}[1]{Refs.~{\cite{#1}}}

\title{Comparing quantumness criteria}

\author{J\'er\^ome Martin\inst{1} \and Amaury Micheli,\inst{2,1} \and
  Vincent Vennin\inst{3,1}} \shortauthor{J. Martin \etal}

\institute{
\inst{1} Institut d'Astrophysique de Paris, CNRS \& Sorbonne Universit\'e, UMR 7095 98
bis boulevard Arago, 75014 Paris, France
\\
\inst{2} Universit\'e Paris-Saclay, CNRS/IN2P3, IJCLab, 91405, Orsay,
France
\\
\inst{3} Laboratoire de Physique de l'Ecole Normale Sup\'erieure,
ENS, CNRS, Universit\'e PSL, Sorbonne Universit\'e, Universit\'e Paris
Cit\'e, F-75005 Paris, France }

\abstract{Measuring the quantumness of a system can be done with a
  variety of methods. In this article we compare different criteria,
  namely quantum discord, Bell inequality violation and
  non-separability, for systems placed in a Gaussian state. When the
  state is pure, these criteria are equivalent, while we find that
  they do not necessarily coincide when decoherence takes
  place. Finally, we prove that these criteria are essentially
  controlled by the semi-minor axis of the ellipse representing the
  state's Wigner function in phase space.}

\begin{document}

\maketitle

%%%%%%%%%%%%%%%%%%%%%%%%
\section{Introduction}

The characterisation of ``classicality'' and ``quantumness'' in
quantum systems has become a topic of major importance in several
branches of modern physics. Indeed, maybe surprisingly, it is not
always trivial to establish whether a system behaves ``classically''
or ``quantum-mechanically''. This question is especially important
when one tries to understand the nature of a physical phenomenon.

For instance, in cosmology, it is well-known that primordial
perturbations are very well
reproduced~\cite{planckcollaborationPlanck2018Results2020a} by vacuum
quantum fluctuations, amplified by gravitational
instability~\cite{Mukhanov:1981xt, Mukhanov:1982nu,
  Starobinsky:1982ee, Guth:1982ec, Hawking:1982cz, Bardeen:1983qw}
during an early epoch of accelerated expansion named
inflation~\cite{Starobinsky:1980te, Sato:1980yn, Guth:1980zm,
  Linde:1981mu, Albrecht:1982wi, Linde:1983gd}.  However, the quantum
origin of those primordial perturbations has never been tested
directly and, in practice, they are mostly treated by astronomers as
classical, stochastic fluctuations. The reason why this is possible is
that, under peculiar circumstances, and for certain observables, a
quantum system can be mimicked by a classical
one~\cite{Lesgourgues:1996jc, Martin:2015qta, Martin:2017zxs}.
However, if a genuine quantum signature could be detected in
cosmological observables, that would shed light on fundamental issues
such as the need to quantise gravitational degrees of freedom or the
emergence of classicality at cosmological
scales~\cite{Sudarsky:2009za, Pinto-Neto:2011fdl,
  martinCosmicInflationQuantum2019, Martin:2019jye,
  micheliQuantumCosmologicalGravitational2022a,Banerjee:2016udy}.

The same need to distinguish classical from quantum processes appears
in analogue gravity, where phenomena involving gravitational physics
are mapped to condensed-matter systems. In these setups, particles can
either be created by quantum channels or by the classical
amplification of a thermal bath~\cite{Busch:2014vda}. The latter
mechanism is always present when conducting experiments at finite
temperature. A quantum test is a way to tell the two populations apart
and to demonstrate the existence of a quantum channel in these
experiments~\cite{jaskulaAcousticAnalogDynamical2012,
  Busch:2014vda,
  steinhauerObservationQuantumHawking2016,
  robertsonAssessingDegreesEntanglement2017}.

In quantum technologies, the distinction between quantum and classical
behaviours is also central, since ``quantumness'' is a crucial
resource \eg in quantum
computing~\cite{jozsaRoleEntanglementQuantum2003} and quantum
cryptography~\cite{ekertQuantumCryptographyBased1991,pironioRandomNumbersCertified2010}.

This has led various notions of ``quantumness'' to be put forward. One possible approach is to consider correlations between sub-parts of
a given system, and to determine whether or not they can be reproduced
by classical random variables. This route gave rise to the celebrated
Bell inequalities
\cite{clauserProposedExperimentTest1969,banaszekQuantumNonlocalityPhase1999,gourOptimizationBellInequality2004},
quantum steering~\cite{wisemanSteeringEntanglementNonlocality2007},
different measures of entanglement
(non-separability~\cite{wernerQuantumStatesEinsteinPodolskyRosen1989},
multipartite entanglement~\cite{horodeckiQuantumEntanglement2009},
entanglement witnesses~\cite{Horodecki:1996nc}, \etc), quantum
discord~\cite{Henderson:2001,Zurek:2001,beraQuantumDiscordIts2018},
\etc.

Another possible approach, leading to a second class of criteria, is
to make use of phase-space formulations of quantum mechanics. For
instance, the non-positivity of the Wigner
function~\cite{Wigner:1932eb} or the absence of the
P-representation~\cite{Sudarshan:1963ts,Glauber:1963tx} have been
viewed as criteria signalling the quantumness of a
system~\cite{2013PhRvA..87f2104G, 2018ApPRv...5d1104W}.

How these different criteria are related is a non-trivial question.
In pure states, it is known that quantum discord reduces to
entanglement entropy~\cite{beraQuantumDiscordIts2018}, which only
vanishes in separable states, and that all non-separable states
violate a Bell
inequality~\cite{wernerQuantumStatesEinsteinPodolskyRosen1989}. For
mixed states however, these relations become more elusive (for
instance non-separability is only a necessary condition for
Bell-inequality
violation~\cite{wernerQuantumStatesEinsteinPodolskyRosen1989}).

In this article, our goal is to investigate the relations between
different criteria in a subclass of quantum states where explicit
calculations can be performed. We want to determine in which cases
they lead to the same conclusion regarding the quantumness of a
system, and in which cases they differ. In practice, we consider two
continuous degrees of freedom placed in two-mode squeezed thermal
states and analyse the link between three quantum criteria:
non-separability, quantum discord and a Bell inequality.

%%%%%%%%%%%%%%%%%%%%%%%%
%\section{2-mode squeezed states}
\section{Gaussian states}

Let us consider two continuous degrees of freedom $q_1$ and $q_2$,
with conjugated momenta $p_1$ and $p_2$, arranged into the phase-space
vector $\hat{\boldmathsymbol{R}}_{1/2} = \left( \hat{q}_{1} ,
\hat{p}_{1} ,\hat{q}_{2} , \hat{p}_{2} \right)^{\mathrm{T}}$ with
$\left[ \hat{q}_{i} , \hat{p}_j \right] = i \delta_{ij}$. Their
quantum state is represented by the density matrix $\hat{\rho}$.  For
a given quantum operator $\hat{O}$, the Weyl transform
\begin{align}
\label{eq:defweyl}
  & \tilde{O}(\boldmathsymbol{R}_{1/2})\equiv \int\mathrm{d}u_1\,
  \mathrm{d}u_2\, 
e^{-ip_1 u_1 - i p_2 u_2}
\nonumber \\ & \times 
\left\langle
  q_1+\frac{u_1}{2},q_2+\frac{u_2}{2}\right\vert \hat{O} \left\vert
  q_1-\frac{u_1}{2},q_2-\frac{u_2}{2}\right\rangle
\end{align}
yields a scalar function in phase space. The Wigner function $W$ is
the Weyl transform of the density matrix~\cite{caseWignerFunctionsWeyl2008},
$W=\tilde{\rho}/(2\pi)^2$, and is such that the expectation value of
any quantum operator $\hat{A}$ is given by the phase-space average of
its Weyl transform against the Wigner function,
\begin{equation}
\label{eq:quantum_averages_as_classical_averages}
\left \langle \hat{A} \right \rangle = \int  \tilde{A}
\left( \boldmathsymbol{R}_{1/2} \right)
W(\boldmathsymbol{R}_{1/2})\, \mathrm{d}^4 \boldmathsymbol{R}_{1/2}  \, .
\end{equation}
This is why the Wigner function is often referred to as a
``quasi-probability'' distribution function.

A Gaussian state is defined as a state whose Wigner function is
Gaussian. All information about the state is then contained in the
covariance matrix
\begin{align}
\label{eq:def:covariance}
\gamma_{ab}=\langle\{\hat{R}_a,\hat{R}_b\}\rangle \, ,
\end{align}
where $\hat{R}_a$ refers to the components of the vector
$\hat{\boldmathsymbol{R}}_{1/2}$,  $\{\hat{A}, \hat{B}\}
= \hat{A} \hat{B} + \hat{B} \hat{A}$ is the anti-commutator
and the Wigner function reads
\begin{align}
\label{eq:gaussian_wigner}
W(\boldmathsymbol{R}_{1/2})=\frac{1}{\pi^2 \sqrt{\det \gamma}}
\exp\left(-\boldmathsymbol{R}^\mathrm{T}_{1/2}
\boldmathsymbol{\gamma}^{-1} \boldmathsymbol{R}_{1/2} \right) \, .
\end{align}

Let us also introduce the purity $p\equiv \tr (\hat{\rho}^2)$, which
determines whether the state is pure ($p=1$) or mixed ($p<1$). For a
Gaussian state, the purity is directly related to the determinant of
the covariance matrix~\cite{adessoContinuousVariableQuantum2014}
\begin{align}
\label{eq:purity_gaussian}
  p=\frac{1}{\sqrt{\det \boldmathsymbol{\gamma} } } \, .
\end{align}

Two-mode squeezed vacuua (TMSV) are Gaussian states whose covariance
matrix depend on two parameters only, $r$ and $\varphi$, respectively
called squeezing amplitude and squeezing angle, and
reads~\cite{BarnettRadmore,Caves:1985zz,Schumaker:1985zz}
\begin{align}
\label{eq:gamma:TMSV}
\boldmathsymbol{\gamma}^{\mathrm{TMSV}}  \equiv
\begin{pmatrix}
\boldmathsymbol{\gamma}^{11} & \boldmathsymbol{\gamma}^{12} \\
\boldmathsymbol{\gamma}^{21} & \boldmathsymbol{\gamma}^{22}
\end{pmatrix}\, ,
\end{align}
with 
\begin{align}
\boldmathsymbol{\gamma}^{11} = \boldmathsymbol{\gamma}^{22}
\equiv \cosh{(2r)} \mathds{1}_{2} \, ,
\end{align}
and
\begin{align}
  \boldmathsymbol{\gamma}^{12} = \boldmathsymbol{\gamma}^{21}
  \equiv -\sinh{2r} 
\begin{pmatrix}
    \cos{2 \varphi}   & \sin{2 \varphi}  \\
    \sin{2 \varphi}  &  -\cos{2 \varphi} 
  \end{pmatrix}\, \, .
\end{align}
TMSV are ubiquitous in modern physics : they appear in quantum
optics~\cite{BarnettRadmore,Caves:1985zz,Schumaker:1985zz}, cold
atoms~\cite{2008Natur.455.1216E,2010PhRvA..82b1806N} as well as in the
study of
inflation~\cite{grishchukSqueezedQuantumStates1990a,grainCanonicalTransformationsSqueezing2020,colasFourmodeSqueezedStates2022,agulloDoesInflationSqueeze2022}
and Hawking radiation~\cite{Hawking:1975vcx, Agullo:2022oye}.  Using
\Eq{eq:purity_gaussian} one can check that they are pure.  In general,
TMSV may become mixed as an effect of decoherence~\cite{Zurek:1981xq,
  Zurek:1982ii, Joos:1984uk}. We will consider the class of two-mode
squeezed thermal states which are defined as Gaussian states with
covariance matrices of the form
\begin{align}
\label{eq:gamma:decohered:TMSV}
\boldmathsymbol{\gamma} =
\frac{\boldmathsymbol{\gamma}^{\mathrm{TMSV}}}{\sqrt{p}}\, ,
\end{align}
where one can check from \Eq{eq:purity_gaussian} that $p$ is indeed
the purity of the state. These states arise for instance for
cosmological perturbations linearly coupled to an environment while
preserving statistical homogeneity
\cite{campoDecoherenceEntropyPrimordial2008a,martinDiscordDecoherence2022},
or when an initial TMSV interacts with two identical independent
thermal baths
\cite{marianDecayGaussianCorrelations2015,ferraroGaussianStatesContinuous2005},
or when the modes are sent through a pure-loss or an additive Gaussian
noise channel~\cite{Genoni:2016qxn}. The two latter channels are
described by simple transformations of the covariance matrix,
respectively given by $ \boldmathsymbol{\gamma} = \eta
\boldmathsymbol{\gamma}^{\mathrm{TMSV}} + \left( 1 - \eta \right)
\mathds{1}_{4}$ where the efficiency parameter $0 \leq \eta \leq 1$
encodes the level of loss/damping experienced across the channel, and
$\boldmathsymbol{\gamma} = \boldmathsymbol{\gamma}^{\mathrm{TMSV}} +
\Delta \mathds{1}_{4} $ where $\Delta \geq 0$ encodes the level of
noise. Both matrices can then be put in the
form~\eqref{eq:gamma:decohered:TMSV}, with effective squeezing and
purity parameters given in \Eqs{eq:effective_squeezing_pure_loss} and
\eqref{eq:effective_squeezing_additive_noise} of the Appendix where
these two channels are studied in details.

In the following we work in terms of these effective squeezing and
purity parameters, such that all setups mentioned above are
encompassed in the analysis. Decoherence is expected to play a key
role in the emergence of classicality, and this simply parameterised
class of states will allow us to study how different criteria respond
to it.

Under a canonical transformation, $\hat{\boldmathsymbol{R}}\to
\boldmathsymbol{T} \hat{\boldmathsymbol{R}}$, where
$\boldmathsymbol{T}$ is a symplectic matrix (\ie it preserves
commutation relations), the covariance matrix changes according to
$\boldmathsymbol{\gamma}\to
\boldmathsymbol{T}\boldmathsymbol{\gamma}\boldmathsymbol{T}^\mathrm{T}$. This
implies that the covariance matrix depends on the set of canonical
variables used to describe a system.

For instance, there exists a partition
$\hat{\boldmathsymbol{R}}_{\mathrm{D}}$ where the covariance matrix is
block diagonal,
\begin{align}
\label{eq:gamma:TMSVS_as_OMSV}
  \gamma^{\mathrm{D}} = \frac{1}{\sqrt{p}}
    \begin{pmatrix}
    \gamma ^{\mathrm{OMSV}} & 0 \\
    0 & \gamma ^{\mathrm{OMSV}}  
  \end{pmatrix}\, ,
\end{align}
with
\begin{align}
\gamma ^{\mathrm{OMSV}}
\equiv
\begin{pmatrix}
    \gamma_{qq} & \gamma _{qp} \\
    \gamma _{pq} & \gamma_{pp}     
  \end{pmatrix}
\end{align}
and
\begin{align}
\label{eq:cov_mat_elements_OMSV}
  \gamma_{qq} &= \left[\cosh(2r) -\cos(2\varphi)\sinh(2r)\right],\\
  \gamma _{pq} & = \gamma _{ \mathrm{qp} }
  = -\sin(2 \varphi) \sinh(2r), \\
\label{eq:cov_mat_elements_OMSVpp}  
  \gamma_{pp} &=  \left[\cosh(2r) +\cos(2\varphi)\sinh(2r)\right],
\end{align}
such that the Wigner function factorises according to $W^{\mathrm{D}}
( \boldmathsymbol{R}^{\mathrm{D}} ) = \bar{W} (q_1^{\mathrm{D}}
,p_1^{\mathrm{D}} ) \bar{W} (q_2^{\mathrm{D}} ,p_2^{\mathrm{D}} )$. In
this basis, the quantum state is nothing but the product of two
identical and uncorrelated one-mode squeezed (thermal) states. If $p=1$ they are one-mode squeezed vacuua (OMSV).

This also implies that quantumness criteria, which characterise the
correlations between two subsystems, obviously depend on the way the
system is partitioned (for instance, the way quantum discord depends
on the choice of partition has been studied
in~\Refs{martinDiscordDecoherence2022,martinQuantumDiscordCosmic2016a}).

\begin{figure}
    \centering
    \onefigure[width=0.48\textwidth]
              {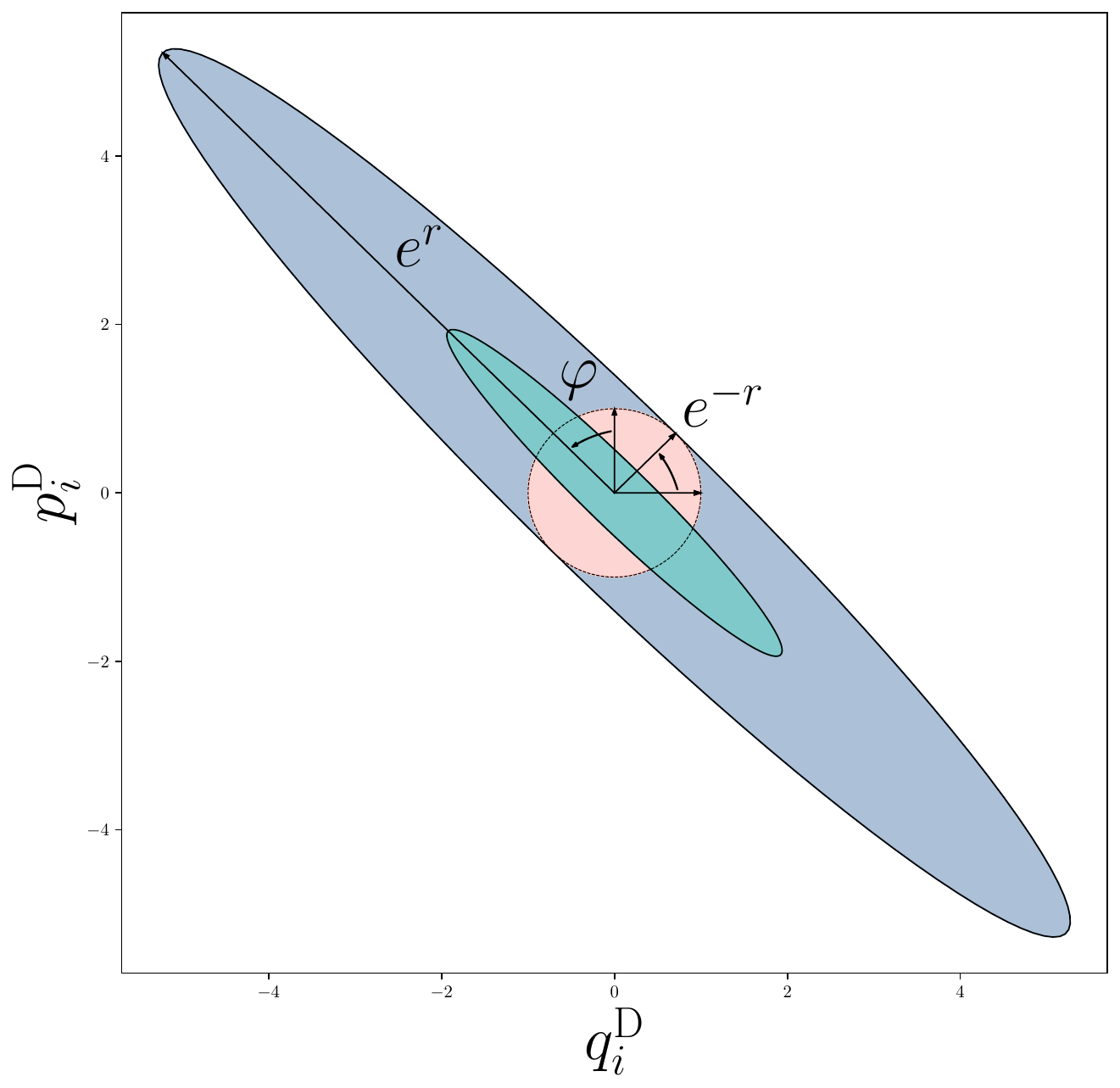}
    \caption{Phase-space $\sqrt{2}$-$\sigma$ contour levels of the
      Wigner function $\bar{W}$. The pink circle corresponds to a
      vacuum state (coherent state) with $p=1$ and vanishing squeezing
      parameter $r=0$. The green ellipse represents a pure state
      $p=1$, slightly squeezed $r = 1$ along the diagonal $\varphi =
      \pi / 4 $. The blue ellipse represents a state with the same
      squeezing parameter $r = 1$ but with purity $p=e^{-4}\approx
      0.018$ such that its semi-minor axis is of the same size as in
      the vacuum state.}
    \label{fig:ellipse_Wigner}
\end{figure}

In practice, their often exists a ``preferred'' basis of operators
corresponding to separately measurable physical degrees of
freedom~\cite{martinRealspaceEntanglementCosmic2021,martinRealspaceEntanglementQuantum2021}
The factorised partition~\eqref{eq:gamma:TMSVS_as_OMSV} is nonetheless
useful as it provides a simple geometric representation of the quantum
state: the contours of $\bar{W}$ are ellipses in the phase space
$(q_i^{\mathrm{D}},p_i^{\mathrm{D}})$, as displayed in
\Fig{fig:ellipse_Wigner}. Their eccentricity is controlled by $r$,
$\varphi$ is the angle between the $q_i^{\mathrm{D}}$-axis and the
semi-minor axis, and the area contained in the ellipses is
proportional to $1/p$.

%%%%%%%%%%%%%%%%%%

\section{Quantumness criteria}

Since the quantum states we consider are fully characterised by the
three parameters $r$, $\varphi$ and $p$, let us express the three
quantumness criteria in terms of these parameters, in order to compare
them.

{\it Quantum Discord.} A first way to characterise the presence of
quantum correlations between two sub-parts of a system is by quantum
discord~\cite{Henderson:2001,Zurek:2001}. The idea is to introduce two
measures of correlation that coincide for classically correlated
setups thanks to Bayes' theorem, but that may differ for quantum
systems. The first measure is the so-called mutual information
$\mathcal{I}$, which is the sum between the von-Neumann entropy of
both reduced sub-systems, minus the entropy of the entire system. The
second measure $\mathcal{J}$ evaluates the difference between the
entropy contained in the first subsystem, and the entropy contained in
that same subsystem when the second subsystem has been measured, where
an extremisation is performed over all possible ways to ``measure''
the second subsystem. $\mathcal{J}$ can be shown to be always less
than $\mathcal{I}$. Quantum discord ${\cal D}$ is defined as the
difference between these two measures and is thus a positive quantity
that only vanishes for classical systems.

For Gaussian states, $\mathcal{I}$ , $\mathcal{J}$ and ${\cal D}$ can
be expressed in terms of the local symplectic invariants of the
covariance
matrix~\cite{adessoContinuousVariableQuantum2014}.\footnote{This means
  that quantum discord is invariant under local symplectic
  transformations, \ie those mixing $q_{i}$ with $p_{i}$ but not with
  $q_j$ and $p_j$. This explains why $\varphi$ does not appear in the
  final expression~\eqref{eq:discord_TMSV}, since it can be changed
  arbitrarily by performing phase-space rotations in each sector.}  It
is shown in~\cite{adessoQuantumClassicalCorrelations2010,giordaGaussianQuantumDiscord2010,pirandolaOptimalityGaussianDiscord2014} that,
for covariance matrices of the form~\eqref{eq:gamma:decohered:TMSV},
quantum discord depends on $r$ and $p$ only and is given by
\begin{align}
\label{eq:discord_TMSV}
{\cal D}(p,r)
 = & f\left[\sigma (p,r) \right]
-2f\left(  p^{-1/2} \right) \, \nonumber  \\
& + f\left[\frac{\sigma (p,r) + p^{-1} }{\sigma (p,r) +1}\right] ,
\end{align}
where the function $f(x)$ is defined for $x\geq 1$ by
\begin{align}
\begin{split}
\label{eq:deff}
  f(x) & \equiv \left(\frac{x+1}{2}\right)
  \log_2\left(\frac{x+1}{2}\right) \\
& -\left(\frac{x-1}{2}\right)
\log_2\left(\frac{x-1}{2}\right),
\end{split}
\end{align}
and 
\begin{align}
  \label{eq:defsigma}
  \sigma(p,r)= \frac{\cosh(2r)}{\sqrt{p}}\, .
\end{align}
Note that in the partition~\eqref{eq:gamma:TMSVS_as_OMSV}, where the
covariance matrix is block-diagonal, the two sub-systems are
uncorrelated hence quantum discord vanishes.

%%%%%%%%%%%%%%%
{\it Bell Inequality.} Another way to characterise the presence of
quantum correlations is via Bell
inequalities~\cite{bellEinsteinPodolskyRosen1964}. When violated, they
allow one to exclude classical and realistic local
theories~\cite{maudlinWhatBellDid2014}. Usually designed for discrete
observables~\cite{clauserProposedExperimentTest1969}(such as spins),
they can also be applied to continuous variables by means of
pseudo-spin
operators~\cite{banaszekQuantumNonlocalityPhase1999,gourOptimizationBellInequality2004}
or via projections on coherent states
\cite{campoInflationarySpectraViolations2006a}.  In this paper we will
use the pseudo-spin operators introduced in
\Refa{gourOptimizationBellInequality2004}
\begin{align}
\label{def:spin_operators}
\hat{\sigma}_x^i = & \int^{ \infty}_{- \infty}
\mathrm{sign}(q_i)\left| q_i \right \rangle\left
\langle q_i \right| \mathrm{d}q_i \, , \\
\hat{\sigma}_y^i = & \, - i \int^{ \infty}_{- \infty}
\mathrm{sign}(q_i)\left| q_i \right \rangle\left \langle
- q_i \right| \mathrm{d}q_i \, , \\
\hat{\sigma}_z^i = & - \int^{ \infty}_{- \infty}
\left| q_i \right \rangle \left \langle - q_i \right| \mathrm{d}q_i \, .
\end{align}
One can check that these operators satisfy the SU(2) commutation
relations
\begin{equation}
  \left[ \hat{\sigma}_{\mu}^i , \hat{\sigma}_{\nu}^j  \right]
  = 2 i \epsilon_{\mu \nu \lambda} \hat{\sigma}_{\lambda}^i \delta^{ij} \, ,
\end{equation}
where $\epsilon_{\mu\nu\lambda}$ is the totally anti-symmetric
tensor.

From these operators we can build a Bell
inequality~\cite{gourOptimizationBellInequality2004,martinObstructionsBellCMB2017}
\begin{equation}
\label{eq:B:GKMR}
\langle \hat{B} \rangle =2\sqrt{\left \langle
  \hat{\sigma}_z^1  \hat{\sigma}_z^2  \right \rangle^2
  +\left  \langle \hat{\sigma}_x^1
  \hat{\sigma}_x^2 \right \rangle^2} \leq 2.
\end{equation}
In order to compute the two-point correlation functions of the
operators $\hat{\sigma}_x$ and $\hat{\sigma}_z$, one can derive their
Weyl transform and make use
of~\Eq{eq:quantum_averages_as_classical_averages}. Since
$\hat{\sigma}_\mu^1$ and $\hat{\sigma}_\mu^2$ act on different degrees
of freedom, the Weyl transform of their product factorises as
\begin{align}
  \widetilde{\sigma_\mu^1 \sigma_\nu^2 }
  =\widetilde{\sigma_\mu^1 }\widetilde{\sigma_\nu^2 } \, ,
\end{align}
and in the appendix we show that
\begin{align}
\label{eq:weyl_spin_z}
  \widetilde{\sigma_z^i }
  =-\pi \delta(q_i )\delta(p_i ), 
  \quad
  \widetilde{\sigma_x^i } =\mathrm{sgn}(q_i)\, ,
   \end{align}
where $\delta$ stands for the Dirac distribution. Together with
\Eq{eq:quantum_averages_as_classical_averages}, this leads to
\begin{align}
\label{eq:zz_correlation}
\left \langle \hat{\sigma}_z^i \hat{\sigma}_z^j \right \rangle
    & =p\, ,\\
\label{eq:xx_correlation}
\left \langle \hat{\sigma}_x^i \hat{\sigma}_x^j \right \rangle 
& =-\frac{2}{\pi}\arcsin \left[ \left| \cos (2\varphi) \right|
  \tanh(2r)\right]\, .
\end{align}
Inserting \Eqs{eq:zz_correlation} and~\eqref{eq:xx_correlation} into
\Eq{eq:B:GKMR} leads to
\begin{align}
\label{eq:Bell_TMSV}
\langle \hat{B} \rangle &=2\sqrt{ p^2 +\frac{4}{\pi^2}
  \arcsin^2 \left[\cos (2\varphi)\tanh(2r)\right]} .
\end{align}
Compared to quantum discord given in \Eq{eq:discord_TMSV}, one can see
that the mean value of the Bell operator $\langle \hat{B} \rangle $
depends on the squeezing angle $\varphi$ in addition to the squeezing
amplitude $r$ and the purity $p$. This is expected since the operators
given in~\Eq{def:spin_operators} are not invariant under local
symplectic transformations.

%%%%%%%
{\it Non-separability.} Finally we consider quantum separability.  A
state is said to be separable in a certain partition if its density
matrix can be written as a statistical mixture of products of density
matrices over the two sub-systems, \ie
\begin{equation}
\hat{\rho} = \sum_{i} \alpha_i \hat{\rho}_{1}^i \bigotimes \hat{\rho}_{2}^i ,
\end{equation}
where $\alpha_i$ are real coefficients. In general, proving that a
state is separable is a non-trivial task, yet, for Gaussian states,
the so-called Peres-Horodecki criterion was proven to be necessary and
sufficient~\cite{simonPeresHorodeckiSeparabilityCriterion2000a}.  In
the appendix we show how to evaluate this criterion for Gaussian
states, in a one-parameter family of partitions that contains
both~\Eq{eq:gamma:decohered:TMSV} and~\Eq{eq:gamma:TMSVS_as_OMSV}. In
the partition corresponding to~\Eq{eq:gamma:TMSVS_as_OMSV}, the state
is, as expected, always separable, while
for~\Eq{eq:gamma:decohered:TMSV} we find that the state is separable
if and only if
\begin{equation}
\label{eq:separability_TMSV}
e^{-2r }  \geq \sqrt{p} \, .
\end{equation}

%%%%%%%%%%%%%%%%%%%%%%%%%%
\section{Results \& Discussion}

\begin{figure*}
\centering
\onefigure[width=0.6\textwidth]{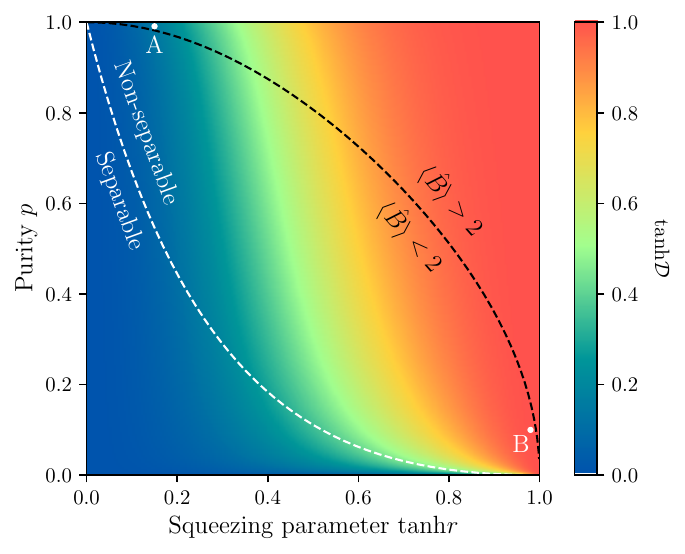}
\caption{Hyperbolic tangent of the quantum discord $\tanh{\cal D}$
  computed from Eq.~(\ref{eq:discord_TMSV}) as a function of the
  purity $p$ and the hyperbolic tangent of the squeezing parameter
  $\tanh{r}$. The dashed white (respectively black) line represents
  the threshold of separability (respectively Bell inequality
  violation) defined as the equality case in \Eq{eq:separability_TMSV}
  [respectively \Eq{eq:Bell_TMSV}].}
\label{fig:purity_and_squeezing}
\end{figure*}

Let us first make connection with the phase-space approaches mentioned
in the introduction. We point out that the thermal squeezed states
considered here always have a \textit{positive} Wigner phase-space
probability distribution, which could make them appear classical. Yet,
as demonstrated, these states can exhibit quantum features. We refer
to
\Refs{ferraroNonclassicalityCriteriaPhaseSpace2012a,revzenBellInequalityViolation2005,martinCosmicInflationQuantum2019}
for detailed discussions of this point. Additionally, for these
states, the absence of a Glauber-Sudarshan P-representation, which is
considered as a sign of non-classicality, is actually equivalent to
the non-separability of the
state~\cite{Kim:1989zza,campoInflationarySpectraPartially2005a} whose
conditions has been computed in \Eq{eq:separability_TMSV}.

We now compare the three different criteria for deciding whether a
system behaves quantum-mechanically or not: the quantum discord
\eqref{eq:discord_TMSV}, the violation of Bell inequalities
\eqref{eq:Bell_TMSV} and the non-separability of the state
\eqref{eq:separability_TMSV}. As mentioned above, the squeezing angle
$\varphi$ can be adjusted by rotating the measurement direction in
phase space. This is why, for the Bell
inequality~\eqref{eq:Bell_TMSV}, which is the only criterion depending
on $\varphi$, we choose to optimise $\varphi$ in order to get the
maximal violation. It corresponds to setting $\varphi=0$. All three
criteria thus depend on $r$ and $p$ only, and are shown
in~\Fig{fig:purity_and_squeezing}. The colour encodes the value of
quantum discord as given by~\Eq{eq:discord_TMSV}, the black dashed
line corresponds to the threshold for Bell-inequality violation,
\Eq{eq:Bell_TMSV}, while the white line stands for the
non-separability criterion as given in~\Eq{eq:separability_TMSV}.

One can check that, for pure state ($p=1$), all criteria are
equivalent: except from the vacuum state ($r=0$), all states have
non-vanishing quantum discord, are non separable and violate the Bell
inequality. In this sense, for a pure Gaussian state, any correlation
is quantum in nature.  For mixed states ($p<1$), non-separability is a
necessary but non-sufficient condition for the Bell-inequality
violation~\cite{wernerQuantumStatesEinsteinPodolskyRosen1989} (\ie the
white line is below the black line), and non-discordant states are
separable~\cite{beraQuantumDiscordIts2018} (\ie the dark blue region
is below the white line).

These results also confirm that decoherence (\ie smaller value for
$p$) is associated to the emergence of classicality. Indeed, for a
given squeezing amplitude $r$, there always exists a value of the
purity parameter $p$ below which the Bell inequality is not violated,
the state is separable and quantum discord is smaller than a given
threshold. The required amount of decoherence (\ie the critical value
for the purity parameter $p$), increases (decreases) with the
squeezing amplitude. This is because, as $r$ increases, the two
subsystems get more entangled, hence it takes more decoherence to
erase quantum features.  In
\cite{marianDecayGaussianCorrelations2015}, the authors had considered
a similar class of states and studied the robustness of
non-classicality measures against decoherence induced by coupling to
thermal baths. In this special case it was also found that the state
becomes classical in the sense that quantum discord asymptotes zero at
large decoherence, and that separability vanishes once decoherence
reaches a certain finite threshold.

\begin{figure*}
\centering
\onefigure[width=0.6\textwidth]{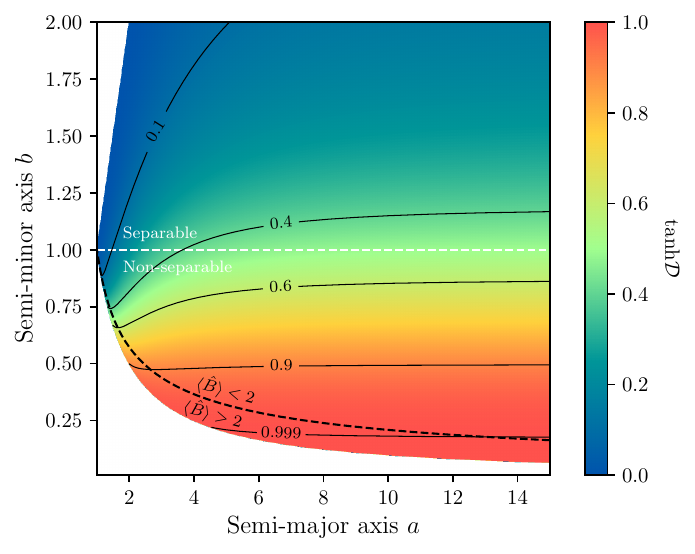} 
\caption{Same criteria as in \Fig{fig:purity_and_squeezing}, as a
  function of the semi-major $a$ and semi-minor $b$ axes of the
  phase-space ellipses depicted in \Fig{fig:ellipse_Wigner}. The solid
  black lines are contour levels of quantum discord. The white region
  corresponds to either $a<b$ or $p>1$, which are both non
  physical. In the large squeezing limit, $b \ll a$, $\langle \hat{B}
  \rangle = 2$ (black dashed line) is equivalent to $b=[ \pi / (8
    a^3)]^{1/5} $.  }
\label{fig:discord_semi_minor_major_axes}
\end{figure*}

Our findings also prompt some reservations about the physical
relevance of the numerical value of quantum discord. Discord is
measured in information bits and, a priori, one may think that it is
an extensive quantity, namely the larger the discord the more
``quantum'' the state. However, one notices
in~\Fig{fig:purity_and_squeezing} that the value of quantum discord at
which the separability or Bell criteria are crossed may be small or
large, depending on the squeezing amplitude. For instance, if the
state is almost pure $p \sim 1$ and the squeezing weak $r \leq 1$,
then one can achieve a non-separable state and/or a Bell inequality
violation while keeping a small quantum discord, see point ``A''; or
for large squeezing and small purity we can both have a large quantum
discord and still satisfy the Bell inequality, see point ``B''. This
suggests that the numerical value of discord itself has no clear
interpretation, at least in this setup and in terms of the other
quantumness criteria.

The behaviour of these three criteria can be further understood in the
phase-space representation. Ignoring the orientation $\varphi$ (which
we have set to its optimal value $\varphi =0$ for Bell inequality
violation), the ellipses of \Fig{fig:ellipse_Wigner} have been
parameterised so far using their area, via $p$, and their
eccentricity, via $r$.  Alternatively, one can describe them by means
of their semi-major, $a$, and semi-minor, $b$, axes, related to $r$
and $p$ by
\begin{equation}
\label{def:semi_major_minor}
 a = e^{r} p^{-1/4}\, , \quad    b = e^{-r} p^{-1/4} \, .
\end{equation}
In particular, we expect $b$, the size of semi-minor axis, to play a
physical role since it encodes the presence or absence of a
sub-fluctuant direction in phase space with respect to the vacuum.

Using Eq.~(\ref{def:semi_major_minor}) all criteria can be expressed
in terms of $a$ and $b$.  The non-separability criterion assumes an
extremely simple form as \Eq{eq:separability_TMSV} is
straightforwardly recast to $b \geq 1$. The fact that the state is
non-separable is then equivalent to the existence of a sub-fluctuant
direction in phase space (for instance, in \Fig{fig:ellipse_Wigner}
the state represented by the green ellipse is non-separable while the
one represented by the blue ellipse is separable).  The expression of
quantum discord and the Bell operator in terms of $a$ and $b$ is not
particularly illuminating but in the large-squeezing and small-purity
limit, \ie $a \gg b \gg 1/a$, in the appendix we show that the discord
also becomes a function of $b$ only (\ie of the sub-fluctuant mode),
namely
\begin{equation}
  \label{eq:largerdiscord}
  {\cal D}(a,b) \rightarrow g \left(1 + 2 b^2\right)
  +\log_2 \left( 1 + \frac{1}{2 b^2} \right) ,
\end{equation}
where $g(x)$ is bounded and defined in \Eq{eq:defg}. 

All criteria are displayed as a function of $a$ and $b$ in
\Fig{fig:discord_semi_minor_major_axes}, where one can check that
$\langle \hat{B}\rangle$ and $\mathcal{D}$ become independent of $a$
in the large-squeezing limit.

\section{Conclusions} In this letter, we compared three
different criteria, quantum discord, Bell inequality violation and
non-separability, aimed at assessing whether a system behaves
quantum-mechanically or not. We have found that, even in a simple
class of Gaussian states, these criteria are inequivalent, \ie a state
can be, at the same time, ``quantum'' according to one criterion and
``classical'' according to another one. However, in the large
squeezing limit these criteria were found to be mainly controlled by
the amplitude of the sub-fluctuant mode.  There is no natural
threshold for the value of quantum discord at which the other two
criteria are crossed, and we found that decoherence always leads to
more classical states regardless of the criterion being used.

This analysis could be extended to non-Gaussian
states~\cite{walschaersNonGaussianQuantumStates2021}, which are known
to behave differently under quantum criteria (for instance, according
to Hudson theorem~\cite{HUDSON1974249} their Wigner functions are
necessarily non-positive if they are pure).

\begin{acknowledgments}
We thank Scott Robertson for interesting discussions and useful
suggestions.  A.~Micheli is supported by the French National Research
Agency under the Grant No. ANR-20-CE47-0001 associated with the
project COSQUA.
\end{acknowledgments}

\bibliographystyle{plain}
\bibliography{biblio}

\onecolumn

\section{Appendix}

In the two first sections of this appendix we present the technical
details relevant for computing the expectation value of the Bell
operator and the separability criterion for Gaussian states. In the
third section, we show that, in the limit $r \gg 1$ and
$p\ll1$, quantum discord is only controlled by $b$, the size of the
semi-minor axis introduced in the main text.  In the last section we analyse the effect of two specific Gaussian noisy channels, namely the pure-loss and the additive-noise channels, on the non-classicality of a TMSV using the criteria identified in the text.

\subsection{Weyl transform of spin operators and expectation value
of Bell operator}

We start by presenting the computation of the Weyl transform of the
spin operators defined in Eq.~(\ref{def:spin_operators}). The Weyl
transform of an operator has been defined in~\Eq{eq:defweyl}. We apply
this formula to $\hat{\sigma}^{i}_{z}$, the spin operator of the
  $i^{\mathrm{rm}}$ subsystem along $z$
\begin{align}
\widetilde{\sigma^{i}_{z}} \left( q_{i} , p_{i} \right)
& = \int^{ \infty}_{- \infty}  e^{i p_{i} y} \left \langle q_{i} + \frac{y}{2}
\right| \left(  - \int^{ \infty}_{- \infty}\left| x \right \rangle
\left \langle - x\right| \mathrm{d} x \right)
\left| q_{i} - \frac{y}{2}   \right \rangle \mathrm{d} y  \\
& = - \int^{ \infty}_{- \infty} e^{i p_{i} y}  \int^{ \infty}_{- \infty}
\delta \left( q_{i} + \frac{y}{2} - x \right)
\delta \left( q_{i} - \frac{y}{2} + x \right) \mathrm{d} x \,
\mathrm{d} y \\
& = - \delta \left(  2 q_{i} \right) \int^{ \infty}_{- \infty}
e^{i p_{i} y} \mathrm{d} y  \\
& = - \pi \delta \left( q_{i} \right) \delta \left( p_{i} \right) \, , 
\end{align}
which is the formula given in \Eq{eq:weyl_spin_z}. This readily
gives \Eq{eq:zz_correlation}. Proceeding similarly for the
spin operator along $x$ we get
\begin{align}
\widetilde{\sigma ^{i}_{x}} \left( q_{i} , p_{i} \right)
& = \int^{ \infty}_{- \infty}  e^{i p_{i} y} \left \langle q_{i}
+ \frac{y}{2}   \right| \left(  - \int^{ \infty}_{- \infty}
\mathrm{sign} \left( x \right) \left| x \right \rangle
\left \langle -x\right| \mathrm{d} x \right)
\left| q_{i} - \frac{y}{2}   \right \rangle \mathrm{d} y  \\
& = \int^{ \infty}_{- \infty} e^{i p_{i} y}  \int^{ \infty}_{- \infty}
\mathrm{sign} \left( x \right)
\delta \left( q_{i} + \frac{y}{2} - x \right)
\delta \left( x - q_{i} + \frac{y}{2} \right) \mathrm{d} x
\, \mathrm{d} y  \\
& =  \int^{ \infty}_{- \infty} e^{i p_{i} y}  \, \mathrm{sign}
\left( q_{i} - \frac{y}{2} \right)  \delta \left( y  \right)
\mathrm{d} y  \\
& = \mathrm{sign} \left( q_{i} \right)  \, ,
\end{align}
which is the (second) formula given in
Eq~(\ref{eq:weyl_spin_z}). Using
Eq.~(\ref{eq:quantum_averages_as_classical_averages}) and the Gaussian
Wigner function~(\ref{eq:gaussian_wigner}), the expectation value of
$\hat{\sigma}^{1}_{z} \hat{\sigma}^{2}_{z}$ can then be obtained as
\begin{align}
\left \langle  \hat{\sigma}^{1}_{x} \hat{\sigma}^{2}_{x} \right \rangle
& = \int  \frac{\mathrm{sign} \left( q_1 \right)
\mathrm{sign} \left( q_2 \right)}{\pi^2 \sqrt{\det \gamma}}
\exp\left(-R^\mathrm{T}_{1/2} \gamma^{-1} R_{1/2} \right)
\mathrm{d} q_1 \, \mathrm{d} p_1 \, \mathrm{d} q_2 \, \mathrm{d} p_2  \\
& = \int \frac{\mathrm{sign} \left( q_1 \right)
\mathrm{sign} \left( q_2 \right)}{\pi \sqrt{ \gamma_{qq} \gamma_{pp} }}
\exp\left[ - \frac{ \left( q_1 + q_2 \right)^2  }{2 \gamma_{qq}}
- \frac{ \left( q_1 - q_2 \right)^2  }{2 \gamma_{pp}}  \right]
\mathrm{d} q_1 \, \mathrm{d} q_2   \\
& = - \frac{2}{\pi} \arctan \left[ \frac{\cos (2\varphi)\sinh(2r)}
  { \sqrt{1 +\sin^2 \left( 2 \varphi \right)
      \sinh \left( 2 r \right) } } \right] \\
&=-\frac{2}{\pi}\arcsin \left[ \left| \cos (2\varphi) \right|
  \tanh(2r)\right]\, ,
\end{align}
where in the second line we have performed the integration over $p_1$
and $p_2$, and in the third line over $q_1$ and $q_2$ after having
inserted the expression of $\gamma_{qq}$ and $\gamma_{pp}$ given
by~Eqs.(\ref{eq:cov_mat_elements_OMSV})
and~(\ref{eq:cov_mat_elements_OMSVpp}). The last result is nothing
but~\Eq{eq:xx_correlation}.

\subsection{Separability criterion}

In this section we derive~\Eq{eq:separability_TMSV} of the main text,
\ie the condition for a Gaussian state to be separable. In the
partition leading to~\Eq{eq:gamma:decohered:TMSV}, the result is
known, see for instance~\cite{campoDecoherenceEntropyPrimordial2008a}.
Here we extend this result to the one-parameter family of partitions
considered in~\Refa{martinDiscordDecoherence2022}: starting from
$\hat{\boldmathsymbol{R}}_{\mathrm{D}}$, it is obtained by performing
the canonical transformation $\hat{\boldmathsymbol{R}}_{\mathrm{D}}
\to S (\theta) \hat{\boldmathsymbol{R}}_{\mathrm{D}}$ where $S
(\theta)$ is the symplectic matrix
\begin{equation}
	S \left( \theta \right) = \begin{pmatrix}
		\cos \theta & 0 & 0 & \sin \theta \\
		0 & \cos \theta & -\sin \theta & 0 \\
		\sin \theta \sin (2\theta) &
                \sin \theta \cos (2\theta) & \cos \theta \cos(2\theta)
		& -\cos \theta \sin(2 \theta) \\
		-\sin \theta \cos(2\theta) & \sin \theta \sin(2\theta) &
		\cos \theta \sin(2\theta) & \cos \theta \cos (2\theta)
	\end{pmatrix}\, .
\end{equation}
This class of partitions is parameterised by the angle $\theta$.  The
partition~\eqref{eq:gamma:decohered:TMSV} corresponds to $\theta = -
\pi / 4$, while the factorised partition, \ie the one leading to
\Eq{eq:gamma:TMSVS_as_OMSV}, corresponds to $\theta = 0$. For
arbitrary $\theta$ the covariance matrix reads
\begin{align}
\label{def:covariance_general_theta}
  \gamma=
  \begin{pmatrix}
    \gamma_A & \gamma_C \\
    \gamma_C & \gamma_B
  \end{pmatrix},
\end{align}
with
\begin{align}
  \label{eq:def:gammaA}
  \gamma_A&=
  \begin{pmatrix}
    \displaystyle
    \gamma_{11} \cos^2 \theta+\gamma_{22}\sin^2 \theta & 
    \displaystyle
    \gamma_{12} \cos(2\theta) \\
    \displaystyle
    \gamma_{12}\cos(2\theta) &
    \displaystyle
    \gamma_{22} \cos^2 \theta+\gamma_{11}\sin^2 \theta
  \end{pmatrix},
  \\
    \label{eq:def:gammaB}
  \gamma_B&=
  \begin{pmatrix}
    \displaystyle
\gamma_B\vert_{11} & \gamma_B\vert_{12} \\
     \displaystyle
     \gamma_B\vert_{21} & \gamma_B\vert_{22}
     \end{pmatrix},
  \\
  \label{eq:def:gammaC}
  \gamma_C&=\left(
  \begin{array}{cccc}
    \displaystyle
    \frac12(\gamma_{11}-\gamma_{22})\sin^2(2\theta)
    +\frac12 \gamma_{12}\sin(4\theta) & & &
    \displaystyle
    -\frac14(\gamma_{11}-\gamma_{22})\sin(4\theta)
    +\gamma_{12}\sin^2(2\theta) \\
    \displaystyle
    -\frac14(\gamma_{11}-\gamma_{22})\sin(4\theta)
    +\gamma_{12}\sin^2(2\theta)& & &
    \displaystyle
    -\frac12(\gamma_{11}-\gamma_{22})\sin^2(2\theta)
    -\frac12 \gamma_{12}\sin(4\theta)
  \end{array}\right),
\end{align}
and where the components of $\gamma_B$ are given by
\begin{align}
\gamma_B\vert_{11}&=
\frac{1}{2}\gamma_{11}+\frac{1}{2}\gamma_{22}+\frac12(\gamma_{11}-\gamma_{22})
\cos(2\theta)\cos(4\theta)-\gamma_{12}\cos(2\theta)\sin(4\theta),
\\
\gamma_B\vert_{12}&=\gamma_B\vert_{21}=
\gamma_{12}\cos(2\theta)\cos(4\theta)
+\frac12(\gamma_{11}-\gamma_{22})\cos(2\theta)\sin(4\theta),
\\
\gamma_B\vert_{22}&=
\frac{1}{2}\gamma_{11}+\frac{1}{2}\gamma_{22}
-\frac12(\gamma_{11}-\gamma_{22})\cos(2\theta)\cos(4\theta)
+\gamma_{12}\cos(2\theta)\sin(4\theta).  
\end{align}

For a general covariance matrix the Peres-Horodecki criterion for
separability can be written as~\cite{SIMON1987223}
\begin{align}
\label{eq:general_separability_criterion}
\det \gamma_A \det \gamma _B +\left( \vert \det \gamma_C\vert -1\right)^2
-\tr \left[\gamma_A J^{(1)}\gamma _C J^{(1)}\gamma _B
J^{(1)}\gamma _C^\mathrm{T} J^{(1)}\right]\geq \det \gamma _A+\det \gamma _B \, ,
\end{align}
where the matrix $J^{(1)}$ is defined by $J^{(1)}\equiv
\begin{pmatrix} 0 & 1 \\ -1 & 0 \end{pmatrix}$. Using the above expressions,
straightforward manipulations lead to
\begin{align}
\det \gamma_A = \det \gamma _B &= 
\frac{1}{p}\left[\cosh^2 (2r)-\cos^2 (2\theta )\sinh^2(2 r)\right] 
%\\ &
= \frac{1}{p}-\det \gamma_C,
\\
\label{eq:det_gamma_C}
\det \gamma _C&=-\frac{1}{p}\sinh^2(2r)\sin^2(2 \theta),
\\
\tr \left[\gamma_A J^{(1)}\gamma _C
  J^{(1)}\gamma _B J^{(1)}\gamma _C^\mathrm{T} J^{(1)}\right]
&= -2 \det \gamma_C \left(\frac{1}{p}-\det \gamma_C\right).
\end{align}
Combining the above results, the general
criterion~\eqref{eq:general_separability_criterion} can be written as
a condition on $\det \left( \gamma_{C} \right)$ only, which is always
negative as can be seen in~\Eq{eq:det_gamma_C}. One obtains
\begin{align}
    \left(\frac{1}{p}-\det \gamma_C\right)^2+\left(\det \gamma_C+1\right)^2
    +2 \det \gamma_C \left(\frac{1}{p}-\det \gamma_C\right)\geq 
    2\left(\frac{1}{p}-\det \gamma _C\right) \, .
\end{align}
Using~\Eq{eq:det_gamma_C} the above reduces to
\begin{align}
    \frac{1}{p^2}-\frac{2}{p}+1+4 \det \gamma _C\geq 0\, .
\end{align}
Using \Eq{eq:det_gamma_C} again, one finds
\begin{align}
\label{eq:theta_separability_criterion}
\left( \frac{1}{\sqrt{p}} - \sqrt{p} \right)^2
\geq 4 \sinh^2(2r)\sin^2(2 \theta) \, .
\end{align}
In the partition leading to~\Eq{eq:gamma:decohered:TMSV}, the above
expression can be evaluated with $\theta = - \pi / 4$, a value for
which the previous formula reduces to
\begin{align}
  \left( \frac{1}{\sqrt{p}} - \sqrt{p} \right)^2
  \geq \left( \frac{1}{e^{-2r } } - e^{-2r} \right)^2 .
\end{align}
Given that both $\sqrt{p}$ and $\ee^{-2r}$ are smaller than one, and
since $y \rightarrow y - 1/y$ is a strictly increasing function, this
finally leads to
\begin{align}
\frac{ e^{-2r }}{\sqrt{p}} \geq 1 \, ,
\end{align}
which corresponds to \Eq{eq:separability_TMSV}.

\subsection{Quantum discord in the large-squeezing limit}

Using Eqs.~(\ref{def:semi_major_minor}), we can re-write the
expression~(\ref{eq:discord_TMSV}) of the quantum discord in terms of
the lengths of the semi-major, $a$, and semi-minor axis,
$b$. Eq.~(\ref{eq:discord_TMSV}) only depends on the quantity
$\sigma$, defined by Eq.~(\ref{eq:defsigma}), and $p$. Therefore, we
need to express these two quantities in terms of $a$ and $b$ and one
obtains
\begin{align}
  \sigma =\frac12 \left(a^2+b^2\right)\, , \quad p=\frac{1}{a^2 b^2}.
\end{align}
Combining Eq.~(\ref{eq:discord_TMSV}) and the two above formula, we
get the following expression for the quantum discord as a function of
$a$ and $b$ only
\begin{align}
\label{eq:discord_semi_minor_squeezing}
{\cal D}(a,b) = f\left[\frac12\left(a^2+b^2\right)\right]
- 2f\left(ab\right)+
f\left(\frac{a^2+b^2+2 a^2 b^2}{a^2+b^2+2}\right)\, .
\end{align}

Under this form the quantum discord is expressed as a sum of terms
which have no definite sign and are unbounded.  In order to see that
${\cal D}$ is only controlled by $b$ in the large-squeezing (\ie $a
\gg b$) and small-purity (\ie $ab=1/\sqrt{p}\gg1$) limit, we rewrite
the above as
\begin{equation}
\label{eq:discord_separation_dominant_terms}
{\cal D}(a,b)=g\left[\frac12\left(a^2+b^2\right)\right]
- 2g\left(ab\right)+
g\left(\frac{a^2+b^2+2 a^2 b^2}{a^2+b^2+2}\right)
+\log_2\left[\frac{(a^2+b^2)(a^2+b^2+2a^2b^2)}{2a^2b^2(a^2+b^2+2)}\right] ,
\end{equation}
where we have defined the function $g(x)$ by 
\begin{equation}
\label{eq:defg}
  g(x) = f(x) - \log_{2} \left( \frac{x}{2} \right) - \frac{1}{\ln 2} \, .
\end{equation}
The function $g(x)$ is defined as the difference between $f(x)$ and
its asymptotic value at large argument. One can check that $g(x)$ is a
negative, strictly increasing function, which is bounded by its limits
$\lim_{x \rightarrow 1^+} g(x) = -1/ \ln 2 + 1 \approx -0.44$ and
$\lim_{x \rightarrow + \infty} g(x) = 0$.  The large-squeezing regime
corresponds to $b \gg a$. Since $a b \geq 1$ for the purity to be
smaller than one, large squeezing requires $a \gg 1$, \ie the
semi-major axis must be much larger than its vacuum value. The first
term in \Eq{eq:discord_separation_dominant_terms} therefore vanishes
in this limit.  In addition, for small purity $a b = 1/\sqrt{p} \gg
1$, the second term vanishes as well.  We are thus left with the last
two terms, which, in this limit, read
\begin{equation}
  \label{eq:largerdiscord}
  {\cal D}(a,b) \rightarrow g \left(1 + 2 b^2\right)
  +\log_2 \left( 1 + \frac{1}{2 b^2} \right) \, .
\end{equation}
Therefore, the value of the discord only depends on the size of the
semi-minor axis $b$ as can be seen in the lower-right corner of
\Fig{fig:discord_semi_minor_major_axes}.  Note that asymptotic
expression behaves as expected in the limit of a large semi-minor
axis, $b \gg 1$, where ${\cal D}$ goes to $0$. In the opposite limit,
namely, $b \ll 1$, the first term goes to a finite value while the
second one vanishes.

\subsection{Pure-loss and additive-noise channels}

\begin{figure*}
	\centering
	\onefigure[width=0.6\textwidth]{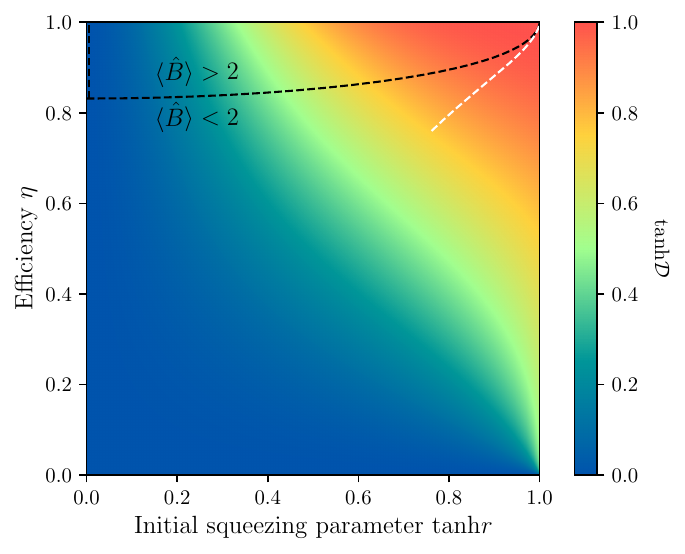}
	\caption{Hyperbolic tangent of the quantum discord $\tanh{\cal
            D}$ computed from Eq.~(\ref{eq:discord_TMSV}) as a
          function of the efficiency $\eta$ and the hyperbolic tangent
          of the initial squeezing parameter $\tanh{r}$. The dashed
          black line represents the threshold of separability Bell
          inequality violation defined as the equality case in
          \Eq{eq:Bell_TMSV}. The vertical piece overlaps and follow
          the line $r=0$ and is represented shifted towards a
          non-vanishing value of $r$ to be visible. The dashed white
          line shows the approximation for the threshold valid for $r
          \gg 1$, plotted for $r\geq1$.  }
	\label{fig:efficiency_and_squeezing}
\end{figure*}

Consider now the effect of a pure-loss channel of efficiency $\eta$ on
a TMSV whose covariance matrix is given by Eq.~(\ref{eq:gamma:TMSV}).
The resulting covariance matrix,
$\eta\boldmathsymbol{\gamma}^{\mathrm{TMSV}} + \left( 1 - \eta \right)
\mathds{1}_{4}$, can be recast in the form of
Eq.~(\ref{eq:gamma:decohered:TMSV}) using the following effective
squeezing parameters and purity
\begin{align}
  \label{eq:effective_squeezing_pure_loss}
  r^{\prime} &= \frac{1}{2} \mathrm{arctanh}
  \left[  \frac{\eta \sinh \left( 2 r\right)}
    { \eta \cosh (2 r) + 1 - \eta }  \right]    ,\quad
  \varphi ^{\prime} = \varphi \, , \quad
  p =  \frac{1}{1 + 4 \sinh^2 \left(  r\right)
    \eta \left( 1 - \eta \right)  }  \, .
\end{align}
We check that, in the limit $\eta\rightarrow 1$ (no loss), the
rescaled squeezing parameters coincide with the original ones and
$p=1$. Using this mapping we can express the quantum discord, the
non-separability and the Bell violation criteria with the help of the
formulas derived in the main text.
We plot these three criteria in
Fig.~\ref{fig:efficiency_and_squeezing}, where we set $\varphi = 0$ to
optimise for the violation of the Bell inequality.

First, we notice that, at fixed value of $r$, the discord increases
with $\eta$, which is intuitive since we expect the quantumness of the
state to be more and more preserved as the loss decreases. Of course,
on the other hand, at fixed efficiency, the discord increases as $r$
increases.

Second, the criterion for
separability~(\ref{eq:general_separability_criterion}), expressed in
terms of $r$ and $\eta$, reads
\begin{equation}
\label{eq:separability_pure_loss}
-16  \eta^2 \sinh^{2} (r) \left[ 1 + \eta \left(2 - \eta\right)
  \sinh^{2} (r) \right] \geq 0 \, ,
\end{equation}
which can only be satisfied if $r=0$. This means that, after having
gone through the loss channel, an initial non-separable state will
always remain non-separable irrespective of its efficiency as expected
for such pure damping~\cite{serafiniEntanglementPurityTwomode2004a}.

Third, in Fig.~\ref{fig:efficiency_and_squeezing}, we have also
represented the threshold for violation of Bell inequality, see the
black dashed line. In terms of $r$ and $\eta$, it is given by the
following expression,
\begin{align}
\label{eq:Bell_pure_loss}
	\langle \hat{B} \rangle & = 2 \sqrt{   \frac{1}{\left[1 + 4 \sinh^2 \left( r\right) \eta \left( 1 - \eta \right)  \right]^2} + \frac{4 }{\pi^2} \arcsin^2 \left[ \frac{\eta \sinh \left( 2 r \right) }{ \eta \cosh (2 r) + 1 - \eta  }  \right] } \, ,
\end{align}
This threshold is discontinuous at $r=0$. Indeed, for $r=0$, the
system is in the vacuum which does not violate the Bell inequality
$\left \langle \hat{B} \right \rangle = 2$. For small but
non-vanishing value of the squeezing parameter, we can expand the
expression of $\langle \hat{B} \rangle$,
\begin{align}
\langle \hat{B} \rangle  \sim  2 + 8 r^2 \eta \left[ \left( \frac{2}{\pi^2} +1 \right) \eta - 1   \right] \, ,
\end{align}
and, as a consequence, the threshold of violation for the Bell
inequality corresponds to $\eta \geq \left(1 + 2 \pi^{-2}
\right)^{-1}\sim 0.83$, which is independent of the squeezing
parameter $r$. We now consider the large $r$ behaviour of the
threshold. The figure shows that for large initial squeezing the level
of loss required to prevent the violation of the Bell inequality
\textit{decreases}. 
This is consistent with the results of~ \cite{jeongDynamicsNonlocalityTwoMode2000} 
where the authors consider a TMSV interacting with two thermal baths, 
and showed that the violation of the Bell inequality 
considered in \cite{banaszekQuantumNonlocalityPhase1999} decreases with the initial squeezing.
Since a large squeezing also implies stronger correlation, and larger value of
the Bell operator initially, this fact might appear surprising at first. 
However, this picture overlooks that the decoherence caused this pure-channel is more efficient for strongly squeezed states. Indeed,  Eq.~\eqref{eq:effective_squeezing_pure_loss} shows that for a channel with fixed efficiency $\eta$, increasing the initial squeezing $r$ of the TMSV will exponentially suppress its purity $p$ after the channel. This decoherence is suppressing the first term in Eq.~\eqref{eq:Bell_pure_loss}, while the stronger correlation increase the second term. The fact that the threshold of Bell inequality violation goes to $\eta=1$ shows that this increase is not sufficient to compensate the decoherence encoded in the first term.
	We can check this behaviour by approximating the curve $\langle \hat{B}\rangle=2$ in
	the vicinity of $r\gg 1$ and $\eta\sim 1$. 
	One finds $\eta\sim
	1-(\pi/8)^{2/5}e^{-6r/5}$, see the white dashed line in
	Fig.~\ref{fig:efficiency_and_squeezing}. This confirms the above
	described behaviour, which illustrates the ``fragility'' of a strongly
	squeezed state.

\begin{figure*}
	\centering
	\onefigure[width=0.6\textwidth]{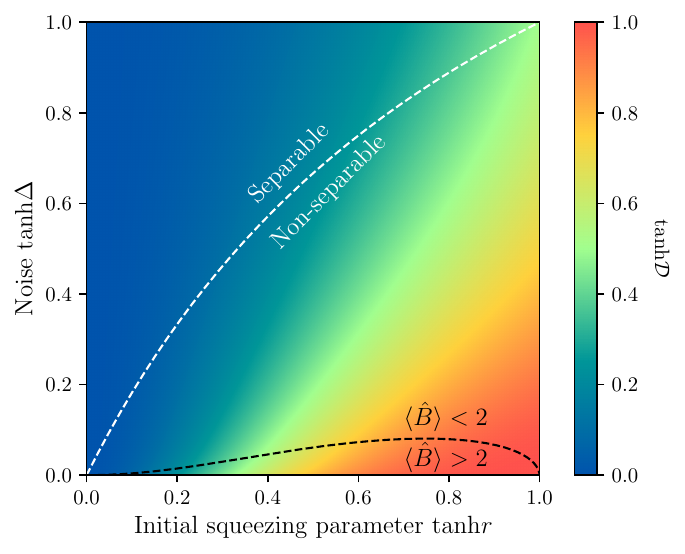}
	\caption{Hyperbolic tangent of the quantum discord $\tanh{\cal D}$
		computed from Eq.~(\ref{eq:discord_TMSV}) as a function of the
		noise $\Delta$ and the hyperbolic tangent of the initial squeezing
		parameter $\tanh{r}$. The dashed white (respectively black) line
		represents the threshold of separability (respectively Bell
		inequality violation) defined as the equality case in
		\Eq{eq:separability_TMSV} [respectively \Eq{eq:Bell_TMSV}].  }
	\label{fig:noise_and_squeezing}
\end{figure*}

Finally, we repeat the same analyses for the additive-noise channel
whose covariance matrix, as already mentioned above, is given by
$\boldmathsymbol{\gamma} = \boldmathsymbol{\gamma}^{\mathrm{TMSV}} +
\Delta \mathds{1}_{4} $, where $\Delta \geq 0$ represents the noise
level. Using the following parameters
\begin{align}
	\label{eq:effective_squeezing_additive_noise}
	r^{\prime} &= \frac{1}{2} \mathrm{arctanh}
        \left[  \frac{ \sinh \left( 2 r\right)}
          { \cosh (2r) + \Delta }  \right]     ,\quad
	\varphi^{\prime} = \varphi  \, , \quad
	p =  \frac{1}{1 + 2 \Delta \cosh (2r) + \Delta^2} \, ,
\end{align}
it can also be put under the form of
Eq.~(\ref{eq:gamma:decohered:TMSV}). Of course, we check that, when
$\Delta \rightarrow 0$, $r'=r$, $\varphi'=\varphi$ and $p=1$. The
exact expressions of the quantum discord, the average value of the
Bell operator and the non-separability threshold can be obtained using
this mapping. Starting from Eq.~(\ref{eq:separability_TMSV}), one can
check that the state is separable if and only if
\begin{equation}
	\Delta \geq 1 - e^{- 2r} \, .
\end{equation}
The expressions of the Bell violation threshold and the quantum
discord can also be derived but are involved and not very
enlightening. We do not reproduce them here. We only want to point out
that a phenomenon similar to that observed in the pure-loss channel
for large initial squeezing also happens for the additive-noise
channel. Namely, as squeezing gets large, the amount of noise required
to destroy the violation of the Bell inequality is reduced. All these
results are summarised in Fig.~\ref{fig:noise_and_squeezing}.

\end{document}